\documentclass[reprint,aps,prl,twocolumn,superscriptaddress]{revtex4-1}
\usepackage[dvips]{graphicx}
\usepackage{dcolumn}
\usepackage{bm}

\begin{document}
\title{Field-Induced Freezing of a Quantum Spin Liquid on the Kagom\'e Lattice}
\author{M. Jeong}
\affiliation{Laboratoire de Physique des Solides, Universit\'e Paris-Sud 11, UMR CNRS 8502, 91405 Orsay, France}
\author{F. Bert}
\affiliation{Laboratoire de Physique des Solides, Universit\'e Paris-Sud 11, UMR CNRS 8502, 91405 Orsay, France}
\email{fabrice.bert@u-psud.fr}
\author{P. Mendels}
\affiliation{Laboratoire de Physique des Solides, Universit\'e Paris-Sud 11, UMR CNRS 8502, 91405 Orsay, France}
\author{F. Duc}
\affiliation{Centre d'\'Elaboration des Mat\'eriaux et d'\'Etudes Structurales, CNRS UPR 8011, 31055 Toulouse, France}
\affiliation{Laboratoire National des Champs Magn\'{e}tiques
Intenses, CNRS UPR 3228, F-31400 Toulouse, France}
\author{J. C. Trombe}
\affiliation{Centre d'\'Elaboration des Mat\'eriaux et d'\'Etudes Structurales, CNRS UPR 8011, 31055 Toulouse, France}
\author{M. A. de Vries}
\affiliation{CSEC and School of Chemistry, University of Edinburgh, Edinburgh, EH9 3JZ, UK}
\affiliation{School of Physics and Astronomy, E.C. Stoner
Laboratory, University of Leeds, Leeds LS2 9JT, UK}
\author{A. Harrison}
\affiliation{CSEC and School of Chemistry, University of Edinburgh, Edinburgh, EH9 3JZ, UK}
\affiliation{Institut Laue-Langevin, 6 rue Jules
Horowitz, F-38042 Grenoble, France}

\begin{abstract}
We report $^{17}$O NMR measurements in the $S=1/2\ (\mathrm{Cu}^{2+})$ kagom\'e antiferromagnet Herbertsmithite $\mathrm{ZnCu_3(OH)_6Cl_2}$ down to 45 mK in magnetic fields ranging from 2~T to 12~T. While
Herbertsmithite displays a gapless spin-liquid behavior in zero
field, we uncover an instability toward a spin-solid phase at sub-kelvin temperature induced by an applied magnetic field. The latter phase shows largely suppressed moments $\lesssim 0.1\mu_\mathrm{B}$ and gapped excitations. The $H-T$ phase diagram suggests the existence of a quantum critical point at the small but finite magnetic field $\mu_0 H_c=1.55(25)$~T. We discuss this finding in light of the perturbative Dzyaloshinskii-Moriya interaction which was theoretically proposed to sustain a quantum critical regime for the quantum kagom\'e Heisenberg antiferromagnet model.

\end{abstract}

\maketitle

Quantum spin liquids (QSL) are appealing states of matter which do not break any symmetry of the spin Hamiltonian. While QSL behaviors are well established for one-dimensional spin systems, their existence in higher dimensions remains questionable. The quantum kagom\'e Heisenberg antiferromagnet (QKHA) is considered as the best candidate to stabilize such a QSL phase in two dimensions~\cite{MisguichL05, Balents10Nat, LhuillierM10}. The specificity of this model relies on the unique combination of strong quantum fluctuations enhanced by low spins $S=1/2$ and high geometrical frustration of the lattice of corner-sharing triangles. Early numerical studies have shown, despite finite size limitations, that the ground state of the QKHA model does not support any simple order parameter \cite{ZengE90PRB,Lecheminant97PRB,Waldtmann98EJPB} and have evidenced an original excitation spectrum with dense sets of low energy excitations
in all spin sectors \cite{Lecheminant97PRB,Waldtmann98EJPB}. A recent state-of-the-art calculation \cite{Yan+10arxiv} favors a QSL ground state with a small gap much like the resonating-valence-bond state \cite{Sachdev92PRB, Anderson73MRB}. Other recent proposals with similar ground-state energies encompass large unit-cell valence-bond-crystals \cite{SinghH07PRB} as well as gapless `critical' spin liquids with algebraic spin correlations \cite{Hastings00PRB,Ran+07PRL}. Clearly, the exact ground state of QKHA  remains a challenging and highly debated issue~\cite{Iqbal11PRB,Lu11PRB}.
In this respect, studying the effect of perturbation, such as disorder, anisotropies or external field, is not only necessary to compare to real materials but also proves to be efficient to discriminate between the competing ground states - critical QSL are expected to be easily destabilized~\cite{Hermele+08PRB,Ran+09PRL} while gapped ones should be more robust.

\begin{figure}
\centering
\includegraphics[width=0.4\textwidth]{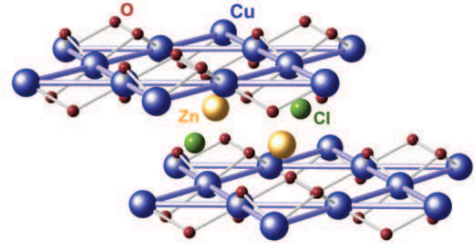}
\caption{Structure of Herbertsmithite. In this ideal structure, Cu and O atoms each occupy one crystallographic site.}
\label{structure}
\end{figure}

To confront theories, a significant step was achieved with the synthesis \cite{Shores+JACS05} of the first ``structurally perfect'' QKHA, $\mathrm{ZnCu_3(OH)_6Cl_2}$, called Herbertsmithite. The electron spin moments of Cu$^{2+}$ ions ($S$=1/2) form undistorted kagom\'e planes,
well separated from each other by diamagnetic Zn$^{2+}$ triangular
planes which ensure the quasi-two-dimensionality of the magnetic
net (Fig.~\ref{structure}). Despite a sizable antiferromagnetic superexchange ${J=180(10)}$~K, Herbertsmithite develops no on-site magnetization \cite{Helton+07PRL, Mendels+07PRL} and remains in an unfrozen, fluctuating state under zero field with short-ranged spin correlations
down to at least
50 mK ($\simeq J/4000$), as expected for a liquid
state \cite{Helton+07PRL,deVries+09PRL}. Inelastic neutron
scattering \cite{Helton+07PRL} and nuclear magnetic
resonance (NMR) \cite{Imai+08PRL,Olariu+08PRL} investigations
consistently point to the absence of a spin gap larger than $\sim
J/200$ and evidence a peculiar power-law temperature dependence
of the spin dynamics, a possible hint of quantum
criticality. At such a low energy scale, small
perturbations to the ideal nearest neighbour Heisenberg model may
become relevant and determine the properties of the ground state.
First, intersite Cu$^{2+}/$Zn$^{2+}$ mixing defects are known to be
present at the level of $5-10$ \% \cite{Helton+07PRL,MendelsB10JPSJ} which could stabilize a gapless valence-bond-glass \cite{Singh10PRL}.
More fundamentally, the lack of inversion
center of the magnetic bonds on the kagom\'e lattice allows for Dzyaloshinskii-Moriya (DM) interaction of spin-orbit origin. In
Herbertsmithite, the DM magnitude is estimated to be $D\sim 0.04J -
0.08 J$ \cite{Zorko+08PRL,ShawishCM10PRB}, not far from a theoretically expected quantum critical
point at $D_c \sim 0.1 J$ \cite{Cepas+08PRB}.

\begin{figure}
\centering
\includegraphics[width=0.5\textwidth]{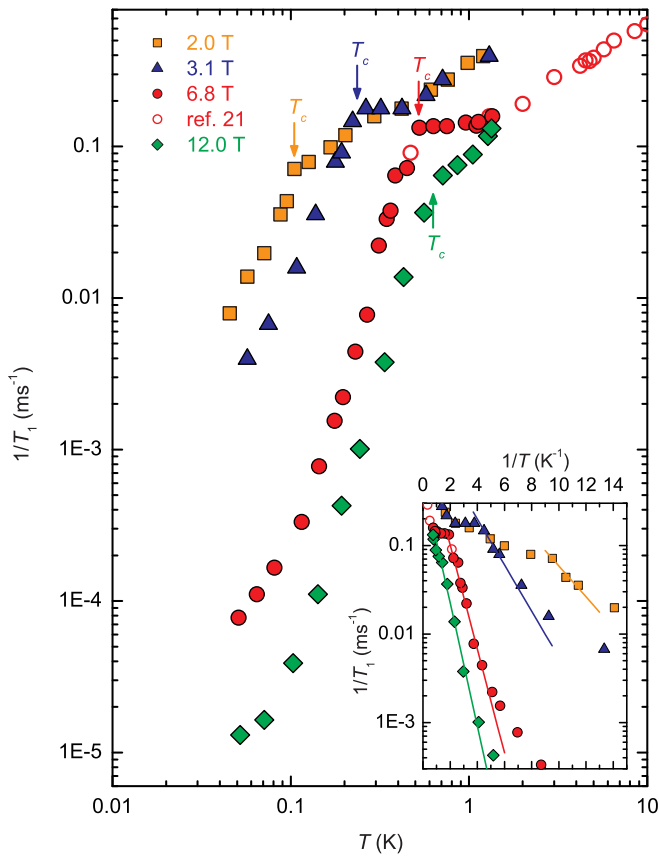}
\caption{$^{17}$O nuclear spin-lattice relaxation rate, $1/T_1$, as a function of  $T$ in Herbertsmithite down to 45 mK. Inset: solid line represents the initial slope below $T_c$. }
\label{relaxation}
\end{figure}

In this Letter, we report the spin dynamics of Hertsmithite in a magnetic field down to very low temperature, $T\simeq J/4000$, probed by $^{17}$O NMR, in order to uncover the ultimate ground-state properties and
test for criticality. The location of oxygen on the superexchange path between two adjacent Cu ions in the kagom\'e lattice confers on the oxygen nucleus a dominant coupling to the magnetic properties of the kagom\'e plane (Fig.~\ref{structure}). This site-selectivity is decisive in the context where the low-$T$ thermodynamic properties and the response of less-coupled local
probes are dominated by interplane defects \cite{Ofer11JPCM,Imai11arXiv}.  We uncover a field induced transition to a frozen phase for applied fields higher than $\mu_0 H_c=1.55(25)$~T and draw the $H-T$ phase diagram.

NMR measurements were performed on the powder sample of
$^{17}$O-enriched Herbertsmithite used in ref.~\cite{Olariu+08PRL}. The spin-echo was obtained with the standard pulse sequence,
$\pi/2-\tau-\pi$. The NMR spectra were
obtained by recording the integrated echo intensity, point-by-point,
while sweeping the external magnetic field with a fixed rf frequency~\cite{SM}.
 The spin-lattice relaxation rate, $1/T_1$, was obtained by saturation-recovery method. The absolute values of $T_1$  were obtained by
scaling the recovery curves to previous results at 6.8~T~\cite{Olariu+08PRL,SM}.

\begin{figure}
\centering
\includegraphics[width=0.5\textwidth]{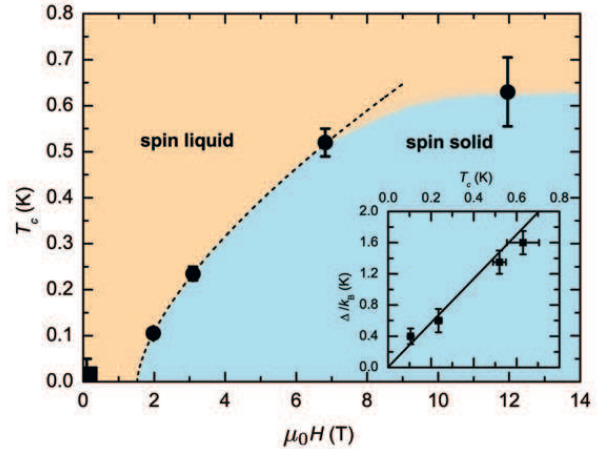}
\caption{$H$-$T$ phase diagram of Herbertsmithite from this study (black circles) and from the absence of spin freezing at least down to 50 mK in $\mu$SR experiment (black square) \protect \cite{Mendels+07PRL}. $T_c$ separates the spin-liquid phase from the spin-solid phase characterized by strongly suppressed fluctuations. The dashed line is a plot of $(H-H_c)^\delta$ with $\mu_0 H_c = 1.53$~T and $\delta = 0.65$ (see text). Inset: $\Delta$ versus $T_c$.}
\label{phasediagram}
\end{figure}

 In Fig.~\ref{relaxation}, we display our
$1/T_1$ results as a function of temperature down to 45 mK in four
different magnetic fields. At
$\mu_0 H=6.8$~T, in contrast to the known modest sublinear $T$-dependence
of $1/T_1\sim T^{0.73}$ above 1~K \cite{Olariu+08PRL}, we find a three orders
of magnitude decrease of $1/T_1$ below $T_c=520(30)$ mK. Both the
marked onset and the magnitude of the drastic slowing down of the
underlying dynamics indicate a transition from the high-$T$
spin-liquid phase to a low-$T$ solid-like phase. This drastic change of the dynamics is detected not only in the vicinity of the peak of the NMR spectrum but also at various positions on the wings, which points to a homogeneous transition. The $1/T_1$ data
taken at 2.0, 3.1 and 12.0 T display a similar drop below a
temperature $T_c$ which is found to increase with the applied field. The corresponding phase diagram is drawn in Fig.~\ref{phasediagram}. By extrapolation to still lower fields, it seems that $T_c$ vanishes at some small but finite critical field $H_c$.
This is in line with the absence of freezing in zero field evidenced
with $\mu$SR~\cite{Mendels+07PRL}. Fitting the data to a phenomenological expression $T_c\propto(H-H_c)^{\delta}$ yields a quantum critical point at $\mu_0 H_c = 1.55(25)$~T and $\delta = 0.6(1)$.
Note that removing the 12~T data point, which may be too far away from the critical regime, improves significantly the quality of the fit (see dashed line in Fig.~\ref{phasediagram}).
 The Zeeman energy
$\mu_\mathrm{B} H_c \sim J/180$, where $\mu_\mathrm{B}$ is the
Bohr magneton, is quite small on the scale of $J$ which underlines the fragility of the zero-field
spin-liquid state of Herbertsmithite.

The inset of Fig.~\ref{relaxation} shows an Arrhenius plot of $1/T_1$ versus
$1/T$. The linear behavior below $T_c$, evident at 6.8~T
and 12.0~T, strongly supports the existence of a gap in the spin
dynamics. This gap in the spin-solid phase likely reveals some magnetic stiffness arising from the DM anisotropic interaction. For $T\lesssim 200$~mK, the relaxation originates from a
residual contribution which is marginal at high fields but sizeable at $\mu_0 H=2.0$ and 3.1~T. This residual relaxation may be attributed to the interplane Cu defects which behave like nearly free $S=1/2$ spins with a small antiferromagnetic coupling of $\sim 1$~K~\cite{MendelsB10JPSJ}. Their contribution is expected to decrease as the field increases, \emph{i.e.} as their magnetization gets saturated. From the initial slope of the $1/T_1$ curves just below $T_c$ (Fig.~\ref{relaxation}, inset), discarding the lowest $T$ data points, we extract the gap $\Delta$. The inset of Fig.~\ref{phasediagram} displays the values obtained for $\Delta$ and their comparison to $T_c$. Within error
bars, $\Delta$ increases linearly with $T_c$ with the slope
$\alpha=\Delta / k_B T_c= 2.3(3)$~\cite{magnon}.

\begin{figure}
\centering
\includegraphics[width=0.43\textwidth]{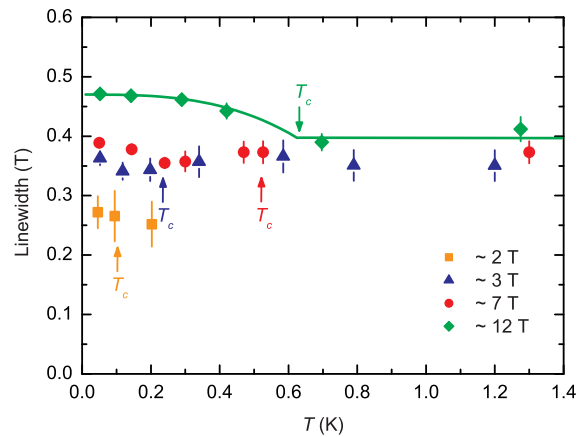}
\caption{Linewidth (full-width
at half-maximum) of the $^{17}$O NMR spectra vs. $T$ for the different fields investigated here, after correction for inhomogeneous spin-spin relaxation effect.}
\label{linewidth}
\end{figure}

\begin{figure}
\centering
\includegraphics[width=0.5\textwidth]{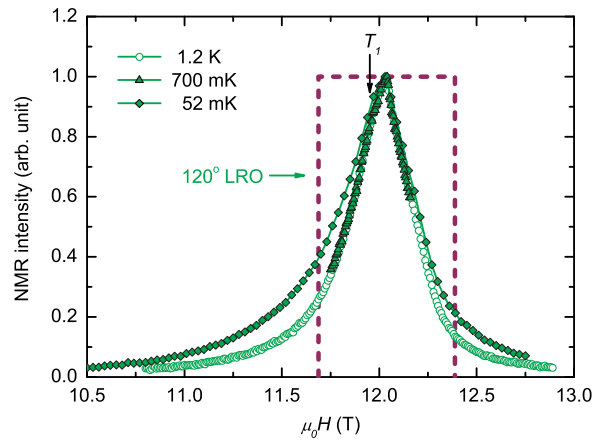}
\caption{$^{17}$O NMR
spectra normalized to their maximum intensity in $\sim$12 T at different $T$. The vertical arrow shows the
location at which $1/T_1$ measurements were performed. The dashed rectangle
represents the expected spectral shape for 120$^{\circ}$
long-range-ordered (LRO) phase with $0.1 \mu_B$ static moments.}
\label{spectra}
\end{figure}

We now turn to the NMR spectra which reflect the distribution of
local magnetization around the probe nuclei. For $H>3$~T and for $T
\gtrsim T_c$, we find that the spectrum width is both field and
temperature independent (see Fig. \ref{linewidth}). This is in agreement with an earlier study
at $\mu_0 H \sim 7$~T \cite{Olariu+08PRL} which demonstrated a sizeable broadening
of the spectra with $1/T$ at high temperature and a saturation below
5~K. It suggests that the width is related to the interplane defects
above $T_c$. The smaller width observed at 2~T and
$T=0.2$~K is in agreement with this picture when the defects are
becoming less saturated. At this smallest field, the width cannot
be measured reliably above $T>0.2$~K because of inhomogeneous,
strong spin-spin relaxation processes. Spectra at 12~T, where the
longer spin-spin relaxations do not affect the lineshape, are
displayed in Fig.~\ref{spectra}. Below $T_c$, the increase of the width, much
like that of an order parameter, contrasts sharply with the absence
of variation observed above $T_c$. This points clearly to
a static magnetization which develops in the kagom\'e plane in the
spin-solid phase. The Gaussian-like lineshape implies a random
static phase whereas, in the case of an ordered phase, the
well-defined internal field at the O site would yield a rectangular
shape for a powder averaged spectrum~\cite{Yamada86}. Given the hyperfine constant
3.5 $\mathrm{T}/\mu_\mathrm{B}$, and independently from any model,
we can extract a typical moment value $\mu_{\rm Cu}\sim 0.1
\mu_{\mathrm B}$, an order of magnitude smaller than the full moment
for a spin 1/2. This small value likely reflects a renormalization
of the moment by strong quantum fluctuations. The absence of any
clear broadening below $T_c$ for the lower fields suggests even
weaker Cu moments, in line with an enhanced quantum reduction in the
vicinity of the critical point $H_c$.

We now aim to discuss which model could best account for a weak-field induced transition in the context of quantum criticality. The algebraic (critical) spin liquid model \cite{Ran+07PRL} captures both the absence of a gap in zero field and the instability under an applied field. A transition to an XY long-range ordered phase and an accompanying on-site magnetization $M \propto H^\alpha$ was predicted to occur at $T_c\propto H$ \cite{Ran+09PRL}. This is not
consistent with our data which suggest a finite critical field.
Also, the presence of a sizeable DM interaction in Herbertsmithite
should induce long-range order already in zero field in this
model \cite{Hermele+08PRB}.

A different approach using exact diagonalization \cite{Cepas+08PRB}
or Schwinger boson techniques \cite{MessioCL10PRB, HuhFS10PRB}
consists, on the contrary, of including the DM interaction in the
Heisenberg model as a starting point. Although the QKHA model may be
gapped~\cite{Yan+10arxiv}, the DM perturbation might naturally account
for the gapless zero-field phase of Herbertsmithite by admixing
singlet and triplet states. Also in this framework, Herbertsmithite
lies close to a quantum critical point that is in fact sustained by
DM anisotropy. A quantum critical region is expected to open at
finite temperatures within a finite range of a control parameter
around its critical value \cite{Sachdev08NatPhys}, which could in
this case be the magnetic field. From the point of view of
symmetry-breaking, the effects of the DM and Zeeman terms may in
essence not be very different. Such an interplay between these two
terms was established for $S=1/2$ Heisenberg antiferromagnetic
chains, a canonical example of a QSL state in one
dimension \cite{OshikawaA07PRL}. Then, Herbertsmithite could be
driven from a QSL to a spin-solid phase by a moderate
field, $H_c \sim D_c-D$. In this framework, the smallness of the moment expresses the
weakness of the order parameter in the vicinity of the quantum
critical point \cite{Cepas+08PRB}. Moreover, the quantum critical
region is characterized by the absence of energy scale other than
temperature which can explain the power-law behavior of spin
correlations \cite{Helton+07PRL, Imai+08PRL, Olariu+08PRL,
deVries+09PRL}. It is then interesting to compare the results with other kagom\'e compounds of different $D/J$, e.g. vesignieite, with possibly $D/J > 0.1$, shows an unusual heterogeneous ground state where spin liquid and frozen moments coexist~\cite{Colman11,Quilliam11}.

Our results share certain features with the field-induced transition recently reported for the organic triangular quantum antiferromagnets $\kappa\mathrm{-(BEDT-TTF)_2Cu_2(CN)_3}$ and $\mathrm{EtMe_3Sb[Pd(dmit)_2]_2}$ which lie close to Mott transition and show a QSL ground state \cite{Pratt+11Nat,Itou+10NatP}. As in Herbertsmithite, small magnetic fields induce the freezing of the zero-field QSL ground state toward a spin-solid phase with strongly suppressed moments. In NMR measurements, no critical divergence in $1/T_1$ nor critical broadening in the spectra are detected \cite{Shimizu03PRL, Shimizu06PRB, Itou+10NatP}, similarly to Herbertsmithite as shown here.  This similarity is quite intriguing since the two lattices, triangular and kagom\'e, have been believed for long to yield very different ground-states and excitation spectra according to numerical studies \cite{Lecheminant97PRB, Waldtmann98EJPB, MisguichL05}.

In conclusion, the QSL ground state of Herbertsmithite appears quite fragile against external magnetic field.
However, in Herbertsmithite, the observed spin freezing likely results from the combination of both a finite applied field and sizeable DM interaction which eventually suggests that the QSL ground state of pure QKHA itself is somewhat robust against perturbations.
It opens up the possibility for a spin gap in the pure QKHA in accordance with most recent calculations \cite{Yan+10arxiv}. The field induced frozen phase that we evidenced here calls for further investigation. Experiments in higher magnetic fields that saturate the interplanar defect contributions and induce larger moments within the kagom\'e plane are desirable. Our finding along with the results reported for the quantum triangular antiferromagnets, prompts a comparative study of the growing class of QSL materials to understand their generic behavior in a broader context of frustration and quantum criticality.

The authors wish to thank O. C\'epas and C. Lhuillier for stimulating and informative discussions. M.J. is grateful to the Embassy of France in Korea for its support through Blaise Pascal Scholarship. This work was supported by the ANR-09-JCJC-0093-01 grant and by the ESF-HFM network.


\begin{thebibliography}{99}
\bibitem{Balents10Nat}L. Balents, Nature {\bf 464,} 199 (2010).
\bibitem{MisguichL05}G. Misguich and C. Lhuillier, in {\it Frustrated Spin Systems}, edited by H. T. Diep (World Scientific, Singapore, 2005).
\bibitem{LhuillierM10}C. Lhuillier, G. Misguich, in {\it Introduction to Frustrated Magnetism}, edited by C. Lacroix, P. Mendels, and F. Mila. (Springer, Berlin Heidelberg, 2010).
\bibitem{ZengE90PRB}C. Zeng and V. Elser, Phys. Rev. B {\bf 42,} 8436 (1990).
\bibitem{Lecheminant97PRB}P. Lecheminant, B. Bernu, C. Lhuillier, L. Pierre and P. Sindzingre, Phys. Rev. B {\bf 56,} 2521 (1997).
\bibitem{Waldtmann98EJPB}C. Waldtmann {\it et al.}, Eur. Phys. J. B {\bf 2,} 501 (1998).
\bibitem{Yan+10arxiv}S. Yan, D. A. Huse and S. R. White, Science {\bf 332,} 1173 (2011).
\bibitem{Sachdev92PRB}S. Sachdev, Phys. Rev. B {\bf 45,} 12377 (1992).
\bibitem{Anderson73MRB}P. W. Anderson, Mat. Res. Bull. {\bf 8,} 153 (1973).
\bibitem{SinghH07PRB}R. R. P. Singh and D. A. Huse, Phys. Rev. B {\bf 76,} 180407(R) (2007).
\bibitem{Hastings00PRB}M. B. Hastings, Phys. Rev. B {\bf 63,} 014413 (2000).
\bibitem{Ran+07PRL}Y. Ran, M. Hermele, P. A. Lee and X.-G. Wen, Phys. Rev. Lett. {\bf 98,} 117205 (2007).
\bibitem{Iqbal11PRB}Y. Iqbal, F. Becca and D. Poilblanc, Phys. Rev. B {\bf 83,} 100404 (2011).
\bibitem{Lu11PRB}Y.M. Lu, Y. Ran and P.A. Lee, Phys. Rev. B {\bf 83,} 224413 (2011).
\bibitem{Hermele+08PRB} M. Hermele, Y. Ran, P. A. Lee, X.-G. Wen, Phys. Rev. B {\bf 77,} 224413 (2008).
\bibitem{Ran+09PRL} Y. Ran, W.-H. Ko, P. A. Lee and X.-G. Wen, Phys. Rev. Lett. {\bf 102,} 047205 (2009).
\bibitem{Shores+JACS05}M. P. Shores, E. A. Nytko, B. M. Bartlett and D. G. Nocera, J. Am. Chem. Soc. {\bf 127,} 13462 (2005).
\bibitem{Helton+07PRL}J. S. Helton {\it et al}., Phys. Rev. Lett. {\bf 98,} 107204 (2007).
\bibitem{Mendels+07PRL}P. Mendels {\it et al}., Phys. Rev. Lett. {\bf 98,} 077204 (2007).
\bibitem{deVries+09PRL}M. A. de Vries {\it et al}., Phys. Rev. Lett. {\bf 103,} 237201 (2009).
\bibitem{Olariu+08PRL}A. Olariu {\it et al}., Phys. Rev. Lett. {\bf 100,} 087202 (2008).
\bibitem{Imai+08PRL}T. Imai, E. A. Nytko, B. M. Bartlett, M. P. Shores and D. G. Nocera, Phys. Rev. Lett. {\bf 100,} 077203 (2008).
\bibitem{MendelsB10JPSJ}P. Mendels and F. Bert, J. Phys. Soc. Jpn. {\bf 79,} 011001 (2010).
\bibitem{Singh10PRL}R. R. P. Singh, Phys. Rev. Lett. {\bf 104,} 177203 (2010).
\bibitem{Zorko+08PRL} A. Zorko {\it et al}., Phys. Rev. Lett. {\bf 101,} 026405 (2008).
\bibitem{ShawishCM10PRB} S. ElShawish, O. C\'epas and S. Miyashita, Phys. Rev. B {\bf 81,} 224421 (2010).
\bibitem{Cepas+08PRB}O. C\'epas, C. M. Fong, P. W. Leung and C. Lhuillier, Phys. Rev. B {\bf 78,} 140405(R) (2008).
\bibitem{Ofer11JPCM} O. Ofer \emph{et al.}, J. Phys.: Condens. Matter \textbf{23}, 164207 (2011).
\bibitem{Imai11arXiv} T. Imai, M. Fu, T.H. Han and Y.S. Lee, Phys. Rev. B \textbf{84}, 020411(R) (2011).
\bibitem{SM} See Supplemental Material at http://link.aps.org/
supplemental/10.1103/PhysRevLett.107.237201 for details on NMR measurements and determination of $T_1$ values.
\bibitem{magnon} The error bar on the slope includes an alternative determination of the gap by fitting the data below $T_c$ to a two magnon process $1/T_1 \propto T \exp(-\Delta / k_B T)$.
\bibitem{Yamada86}Y. Yamada and A. Sakata, J. Phys. Soc. Jpn. \textbf{55}, 1751 (1986)
\bibitem{MessioCL10PRB} L. Messio, O. C\'epas and C. Lhuillier, Phys. Rev. B {\bf 81,} 064428 (2010).
\bibitem{HuhFS10PRB} Y. Huh, L. Fritz and S. Sachdev, Phys. Rev. B {\bf 81,} 144432 (2010).
\bibitem{Sachdev08NatPhys} S. Sachdev, Nat. Phys. {\bf 4,} 173 (2008).
\bibitem{OshikawaA07PRL} M. Oshikawa and I. Affleck, Phys. Rev. Lett. {\bf 79,} 2883 (1997).
\bibitem{Colman11} R. Colman {\it et al.}, Phys. Rev. B {\bf 83,} 180416(R) (2011).
\bibitem{Quilliam11} J. A. Quilliam {\it et al.}, arXiv:1105.4338 (2011).
\bibitem{Pratt+11Nat} F. L. Pratt {\it et al.}, Nature {\bf 471,} 612 (2011).
\bibitem{Shimizu03PRL} Y. Shimizu {\it et al.}, Phys. Rev. Lett. {\bf 91,} 107001 (2003).
\bibitem{Shimizu06PRB} Y. Shimizu {\it et al.}, Phys. Rev. B {\bf 73,} 140407(R) (2006).
\bibitem{Itou+10NatP} T. Itou, A. Oyamada, S. Maegawa and R. Kato, Nat. Phys. {\bf 6,} 673 (2010).
\end{thebibliography}
\end{document}